\documentclass[a4paper,11pt]{article}
\usepackage{pos}

\usepackage{graphicx}
\graphicspath{{images/}}

\usepackage{float}
\usepackage{subfig}
\usepackage{multicol}

\usepackage{hyperref}

\title{Development of the Double Cascade Reconstruction Techniques in the Baikal-GVD Neutrino Telescope}
 \ShortTitle{Double Cascade Reconstruction}

\author[a]{V.A.~Allakhverdyan}
\author[b]{A.D.~Avrorin}
\author[b]{A.V.~Avrorin}
\author[b]{V.M.~Aynutdinov}
\author[c]{R.~Bannasch}
\author[d]{Z.~Barda\v{c}ov\'{a}}
\author[a]{I.A.~Belolaptikov}
\author[a]{I.V.~Borina}
\author[a,1]{V.B.~Brudanin}
\author[e]{N.M.~Budnev}
\author[a]{V.Y.~Dik}
\author[b]{G.V.~Domogatsky}
\author[b]{A.A.~Doroshenko}
\author[a,d]{R.~Dvornick\'{y}}
\author[e]{A.N.~Dyachok}
\author[b]{Zh.-A.M.~Dzhilkibaev}
\author*[d]{E.~Eckerov\'{a}}
\author[a]{T.V.~Elzhov}
\author[f]{L.~Fajt}
\author[g,1]{S.V.~Fialkovski}
\author[e]{A.R.~Gafarov}
\author[b]{K.V.~Golubkov}
\author[a]{N.S.~Gorshkov}
\author[e]{T.I.~Gress}
\author[a]{M.S.~Katulin}
\author[c]{K.G.~Kebkal}
\author[c]{O.G.~Kebkal}
\author[a]{E.V.~Khramov}
\author[a]{M.M.~Kolbin}
\author[a]{K.V.~Konischev}
\author[h]{K.A.~Kopa\'{n}ski}
\author[a]{A.V.~Korobchenko}
\author[b]{A.P.~Koshechkin}
\author[i]{V.A.~Kozhin}
\author[a]{M.V.~Kruglov}
\author[b]{M.K.~Kryukov}
\author[g]{V.F.~Kulepov}
\author[h]{Pa.~Malecki}
\author[a]{Y.M.~Malyshkin}
\author[b]{M.B.~Milenin}
\author[e]{R.R.~Mirgazov}
\author[a]{D.V.~Naumov}
\author[a]{V.~Nazari}
\author[h]{W.~Noga}
\author[b]{D.P.~Petukhov}
\author[a]{E.N.~Pliskovsky}
\author[j]{M.I.~Rozanov}
\author[a]{V.D.~Rushay}
\author[e]{E.V.~Ryabov}
\author[b]{G.B.~Safronov}
\author[a]{B.A.~Shaybonov}
\author[b]{M.D.~Shelepov}
\author[a,d,f]{F.~\v{S}imkovic}
\author[a]{A.E. Sirenko}
\author[i]{A.V.~Skurikhin}
\author[a]{A.G.~Solovjev}
\author[a]{M.N.~Sorokovikov}
\author[f]{I.~\v{S}tekl}
\author[b]{A.P.~Stromakov}
\author[a]{E.O.~Sushenok}
\author[b]{O.V.~Suvorova}
\author[e]{V.A.~Tabolenko}
\author[e]{B.A.~Tarashansky}
\author[a]{Y.V.~Yablokova}
\author[c]{S.A.~Yakovlev}
\author[b]{D.N.~Zaborov}

\affiliation[a]{Joint Institute for Nuclear Research, Dubna, Russia}
\affiliation[b]{Institute for Nuclear Research, Russian Academy of Sciences, Moscow, Russia}
\affiliation[c]{EvoLogics GmbH, Berlin, Germany}
\affiliation[d]{Comenius University, Bratislava, Slovakia}
\affiliation[e]{Irkutsk State University, Irkutsk, Russia}
\affiliation[f]{Czech Technical University in Prague, Prague, Czech Republic}
\affiliation[g]{Nizhny Novgorod State Technical University, Nizhny Novgorod, Russia}
\affiliation[h]{Institute of Nuclear Physics of Polish Academy of Sciences (IFJ~PAN), Krak\'{o}w, Poland}
\affiliation[i]{Skobeltsyn Institute of Nuclear Physics, Moscow State University, Moscow, Russia}
\affiliation[j]{St.~Petersburg State Marine Technical University, St.Petersburg, Russia}

\note{Deceased.}

\emailAdd{eliska.eckerova@fmph.uniba.sk}

\abstract{ The Baikal-GVD is a neutrino telescope under construction in Lake Baikal. The main goal of the Baikal-GVD is to observe neutrinos via detecting the Cherenkov radiation of the secondary charged particles originating in the interactions of neutrinos. In 2021, the installation works concluded with 2\,304 optical modules installed in the lake resulting in effective volume $\sim$ $0.4~\text{km}^{3}$. In this paper, the first steps in the development of double cascade reconstruction techniques are presented.
}

\FullConference{37$^{\rm{th}}$ International Cosmic Ray Conference (ICRC 2021)\\
		July 12th -- 23rd, 2021\\
		Online -- Berlin, Germany}

\begin{document}
\maketitle

\section{Introduction}
The Baikal Gigaton Volume Detector (Baikal-GVD)~\cite{baikal_gvd} is a cubic-kilometer scale neutrino telescope located in the deepest freshwater lake in the world -- Lake Baikal installed approximately 3 - 4~km from shore at depths of $\sim$ 750 - 1275~m. This three dimensional array of photomultiplier tubes aims to detect the Cherenkov radiation emitted in water by products of neutrino interactions.

\begin{figure}[h!]
	\centering
	\begin{multicols}{2}
		\centering
		\subfloat[]{
			\includegraphics[width=0.83\linewidth]{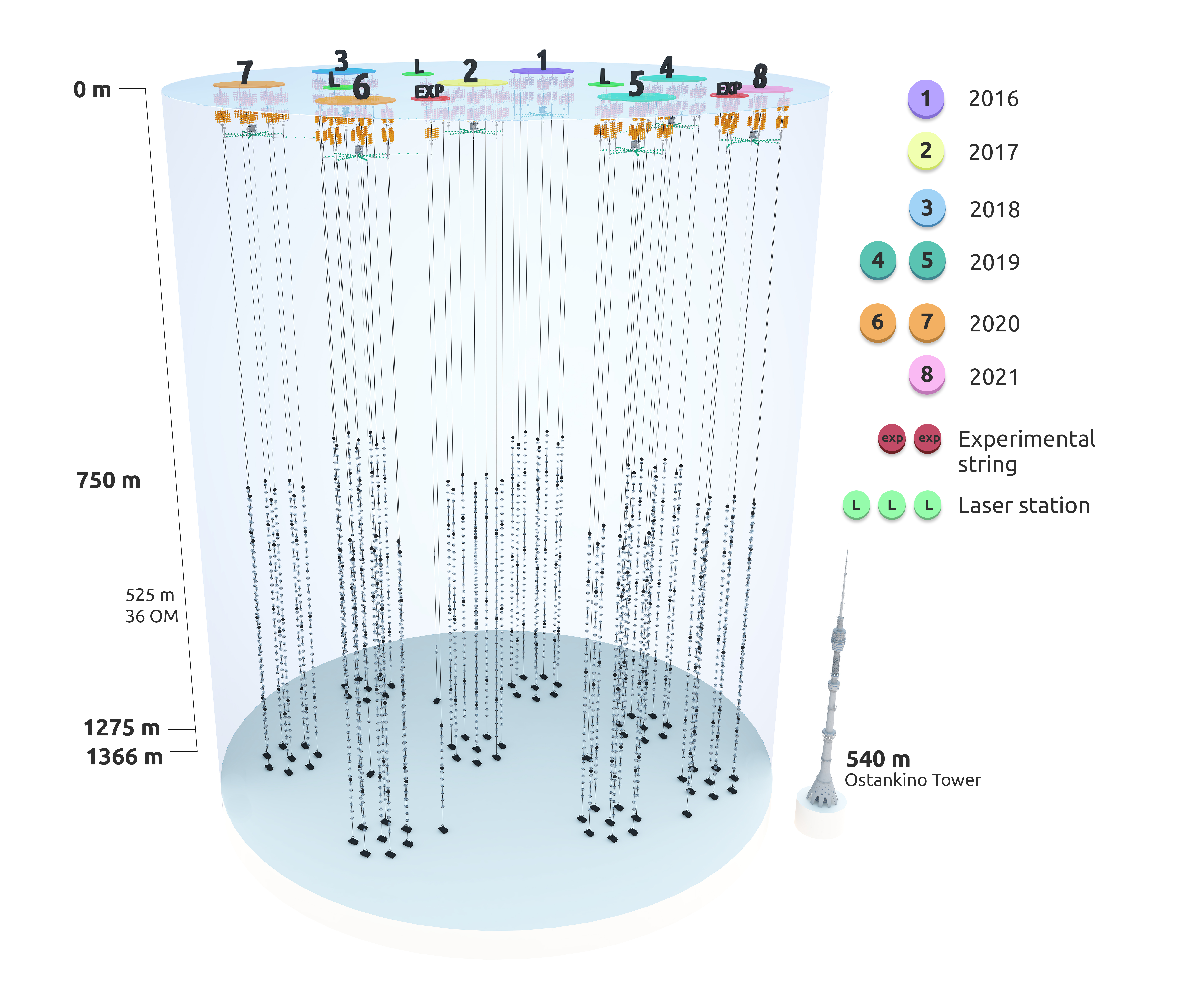}
			\label{obr:design}
		}
		\setlength{\columnsep}{-15.5cm}
		\subfloat[]{
			\includegraphics[width=0.1655\linewidth, trim={0cm 0 0cm -0.3cm},clip]{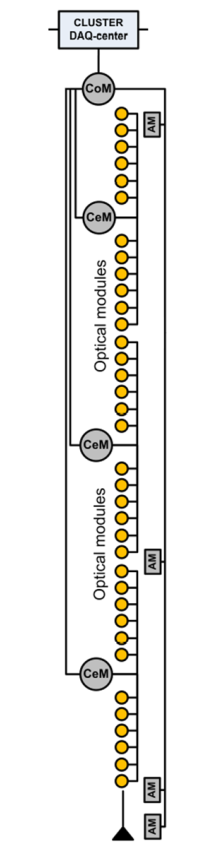}
			\label{obr:string}
		}        
	\end{multicols}
	\caption{a) Design of the Baikal-GVD telescope in 2021. There are 8 clusters, 3 laser strings, and 2 experimental strings installed and operating. b) Illustration of the string of the Baikal-GVD telescope consisting of 36 OMs arranged in 3 sections.}
	\label{fig:Baikal_GVD}
\end{figure}

The basic and independently working unit of the Baikal-GVD detector is called cluster, which consists of 8 strings - one central and seven peripheral separated by about 60 m. The distance between the centers of two neighboring clusters is approximately 300 m (see Fig.~\ref{obr:design}). On every string, 36 Optical Modules (OM) are installed (see Fig.~\ref{obr:string}) with vertical spacing 15~m, resulting in 288 OMs for every cluster. The main component of the OM is a photomultiplier tube with hemispherical photocatode enclosed in pressure-resistant glass sphere with diameter of 42~cm.

One of the methods for astrophysical neutrino detection is an observation of $\nu_{\tau}$ because the rate of $\nu_{\tau}$ production in the atmosphere is almost negligible~\cite{Palladino:Importance_tau}. Therefore observed tau neutrino immediately confirms its astrophysical origin. In the charged current interaction of $\nu_{\tau}$, $\tau$ lepton is created in the hadronic cascade. If $\tau$ decays into electron or hadrons, the second cascade is produced. As a result, the double cascade signature is created.

In this paper, techniques for the reconstruction of double cascades are presented. The first method is a double pulse detection technique for the reconstruction of double cascade events with small distance between cascade vertices. The second method is based on the identification of two distinct cascades created in a single cluster and the third one combines single cluster single cascade reconstruction technique with multi-cluster events studies.

\section{Search for Double Cascades with Double Pulse Detection Method}

To be able to reconstruct double cascade events with very small distance between cascade vertices, the events have to be studied on the waveform level. Thus, double pulse detection method was developed.

\begin{figure}[h!]
	\centering
	\begin{multicols}{2}
		\centering
		\subfloat[]{
			\includegraphics[width=0.85\linewidth, height=5cm]{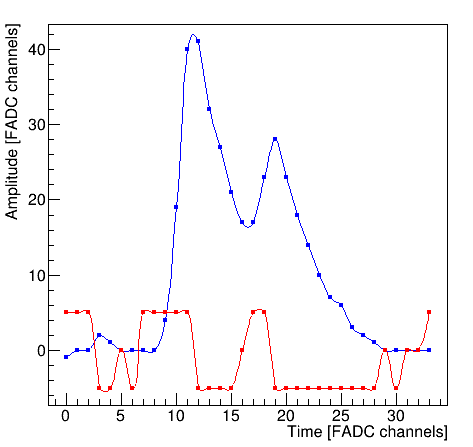}
			\label{obr:diff_method}
		}
		\subfloat[]{
			\includegraphics[width=1.15\linewidth, height=5.3cm]{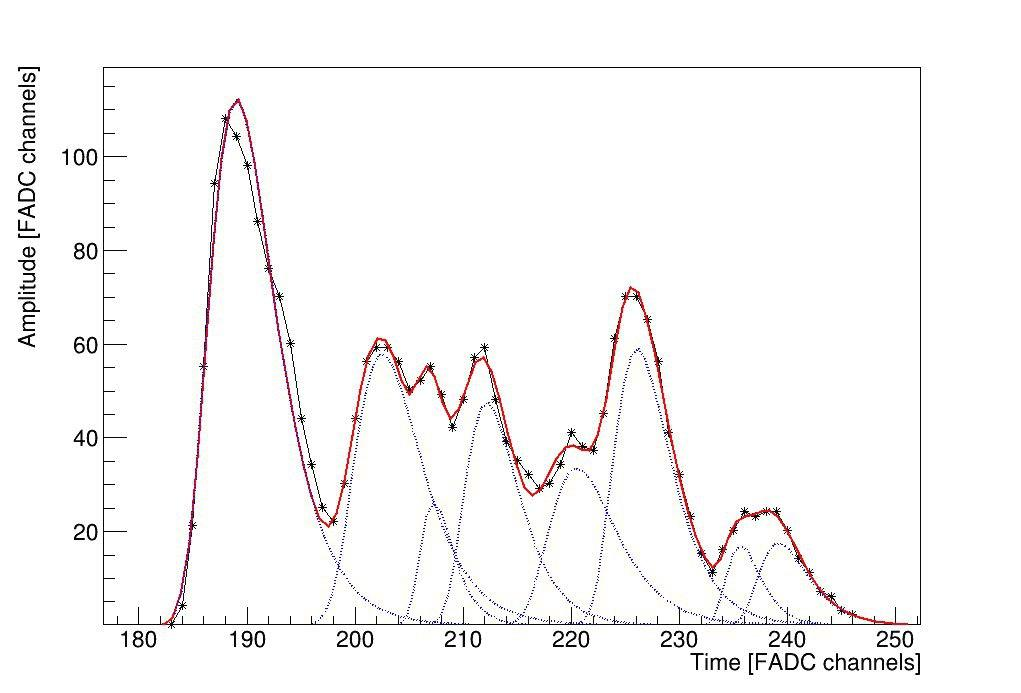}
			\label{obr:Multiple_pulse}
		}        
	\end{multicols}
	\caption{a) Illustration of the differentiation method. The waveform is displayed with the blue line. The red line corresponds to the sign of the first derivative multiplied by a factor of 5 for the sake of visualization. b) Multiple pulse fitting. The black line corresponds to experimental waveform. The fit with sum of the Gumbel functions is displayed with the red line. The black dotted peaks correspond to individual Gumbel functions.}
	\label{fig:Waveforms}
\end{figure}

To tag the potential double pulses, the differentiation method was selected. A potential double pulse is tagged if a sign of the first derivative changes from negative to positive (see Fig.~\ref{obr:diff_method}). Hence a creation of local minimum in the waveform is required.

To determine the minimal time delay between pulses needed for the formation of a local minimum with respect to the charges of both pulses, simulations of the Gumbel functions were studied. The analytical form of the Gumbel function is following:
\begin{equation}
f(x)=a e^{-((\frac{x-\mu}{\beta})+e^{-(\frac{x-\mu}{\beta})})},
\label{eq:Gumbel}
\end{equation}
where $a$ is a normalization coefficient, $\mu$ is a position of maximum, and $\beta$ is a width of the peak. This analysis showed that the minimal time delay between pulses required for formation of a local minimum is $\sim$ 25~ns.

However, many potential double pulses tagged by the differentiation method are fake double pulses created by pedestal oscillations. To suppress them, a multivariate machine learning technique -- Boosted Decision Trees (BDT) from ROOT TMVA package~\cite{Hoecker:TMVA:2007} was implemented. For BDT training, 12~parameters were used. Both datasets, the signal dataset (real double pulses) and background dataset (fake double pulses), contain attributes of approximately 3000 waveforms for training and testing of BDT.

\begin{figure}[h!]
	\centering
	\begin{multicols}{2}
		\centering
		\subfloat[]{
			\includegraphics[width=\linewidth, height=5cm]{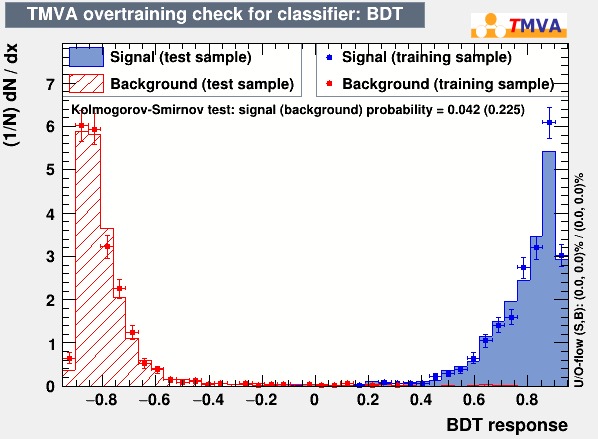}
			\label{obr:BDT_response_distribution}
		}
		\subfloat[]{
			\includegraphics[width=\linewidth, height=5cm]{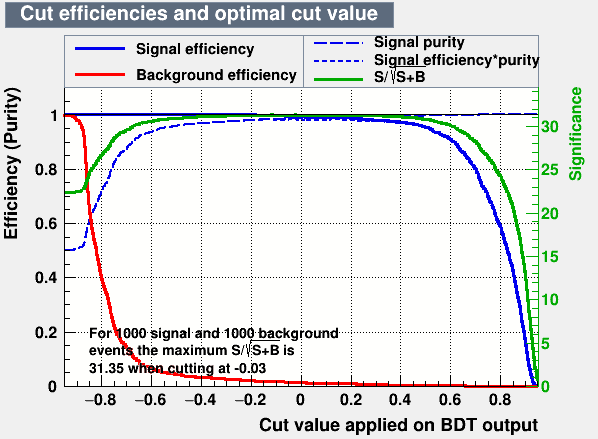}
			\label{obr:BDT_eff_pur}
		}        
	\end{multicols}
	\caption{a) Distribution of BDT response value for background (red color) and signal (blue color). b) Distributions of signal efficiency and background efficiency with respect to the cut value applied on BDT output.~\cite{Eckerova:Thesis:2020} }
	\label{fig:BDT_results}
\end{figure}

The output from trained BDT -- the distribution of BDT response value is shown in Fig.~\ref{obr:BDT_response_distribution}. With a cut on BDT response value at -0.03, the signal efficiency is 99.6\%, and the background efficiency is 1.4\% (see Fig.~\ref{obr:BDT_eff_pur})~\cite{Eckerova:Thesis:2020}.

This method can be used to identify not only the double pulses but any higher multiplicity of pulses (multiple pulses). Conclusively, individual pulses can be fit with the Gumbel function and their parameters can be determined. The demonstration of this procedure applied on experimental pulse is shown in Fig.~\ref{obr:Multiple_pulse}.

\section{Single Cluster Double Cascade Reconstruction Algorithm}

The double cascade reconstruction algorithm consists of three main parts -- signal/noise hits selection, division of pulses for separate reconstruction of two cascades, and two independent single cascade reconstructions. Input to this algorithm is a set of pulses recorded by photomultiplier tubes.

The first step is to separate signal and noise pulses. Pulse with the highest charge is chosen as a reference pulse. Other pulses are selected with respect to the reference pulse according to the causality criterion:
\begin{equation}
\mid t-t_{i} \mid <d_{i}/ v + \delta t,
\label{eq:Causality}
\end{equation}
where $t$ is detection time of the reference pulse, $t_{i}$ is detection time of the pulse to be analyzed, $d_{i}$ is distance between OMs of the two pulses, and $\delta t$ is parameter that determine stringency of causality criterion. There are also additional criteria on neighboring OMs that detected hits. With $\delta t$ = 80~ns, average efficiency of signal pulse selection is at the level of 80\% and purity is 99\%.

In the second step, the set of pulses from the first step is divided to two subsets for two independent single cascade reconstructions. The pulses are selected based on estimated positions and times of cascades.

\begin{figure}[h!]
	\centering
	\begin{multicols}{2}
		\centering
		\hspace{0.7cm}
		\subfloat[]{
			\centering
			
			\includegraphics[width=0.78\linewidth, trim={0cm 3cm 0cm 4cm},clip]{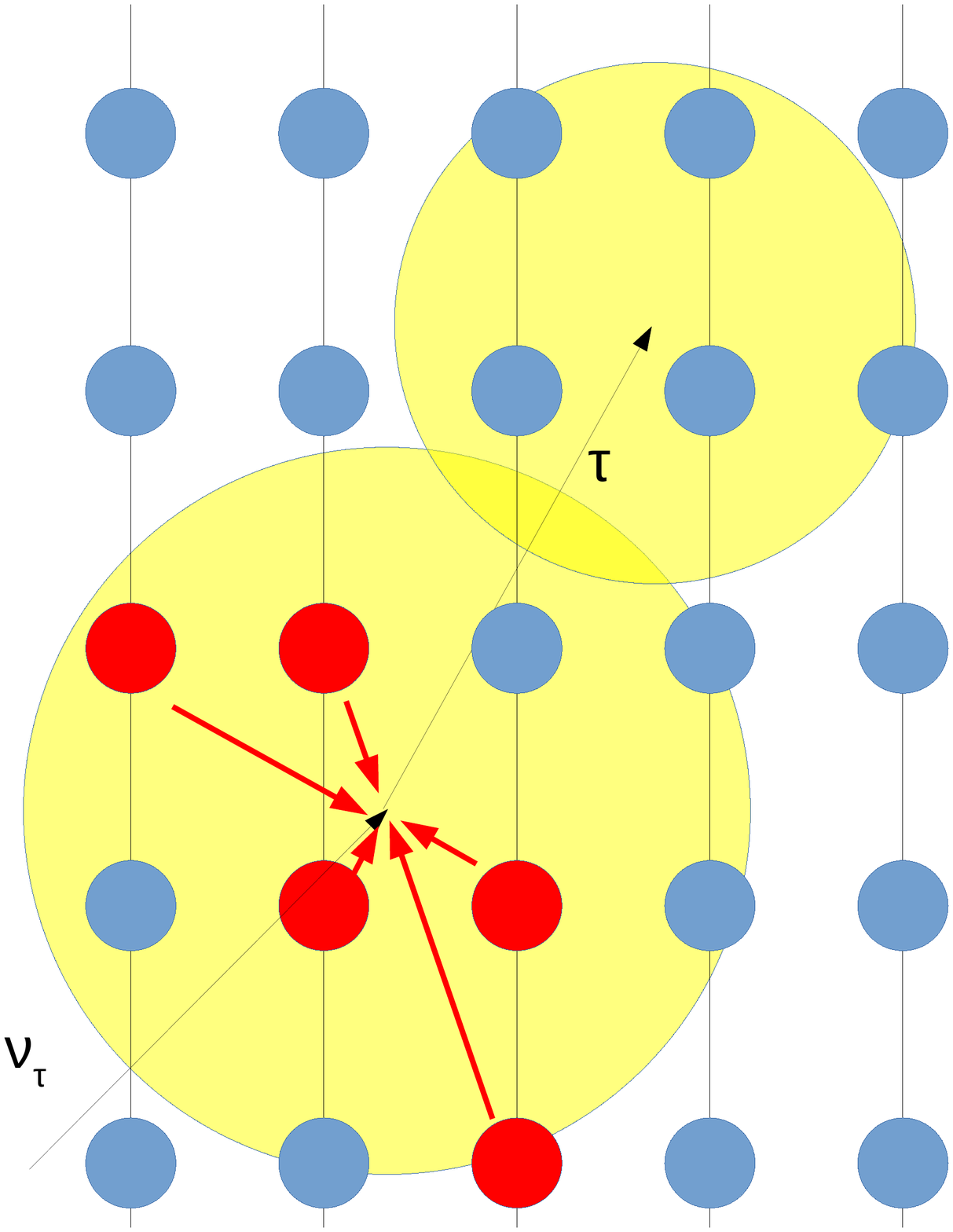} 
			\label{obr:set_5_one_cascade}
		}
	\hspace{0.7cm}
		\subfloat[]{
			\centering
			
			\includegraphics[width=0.78\linewidth, trim={0cm 3cm 0cm 4cm},clip]{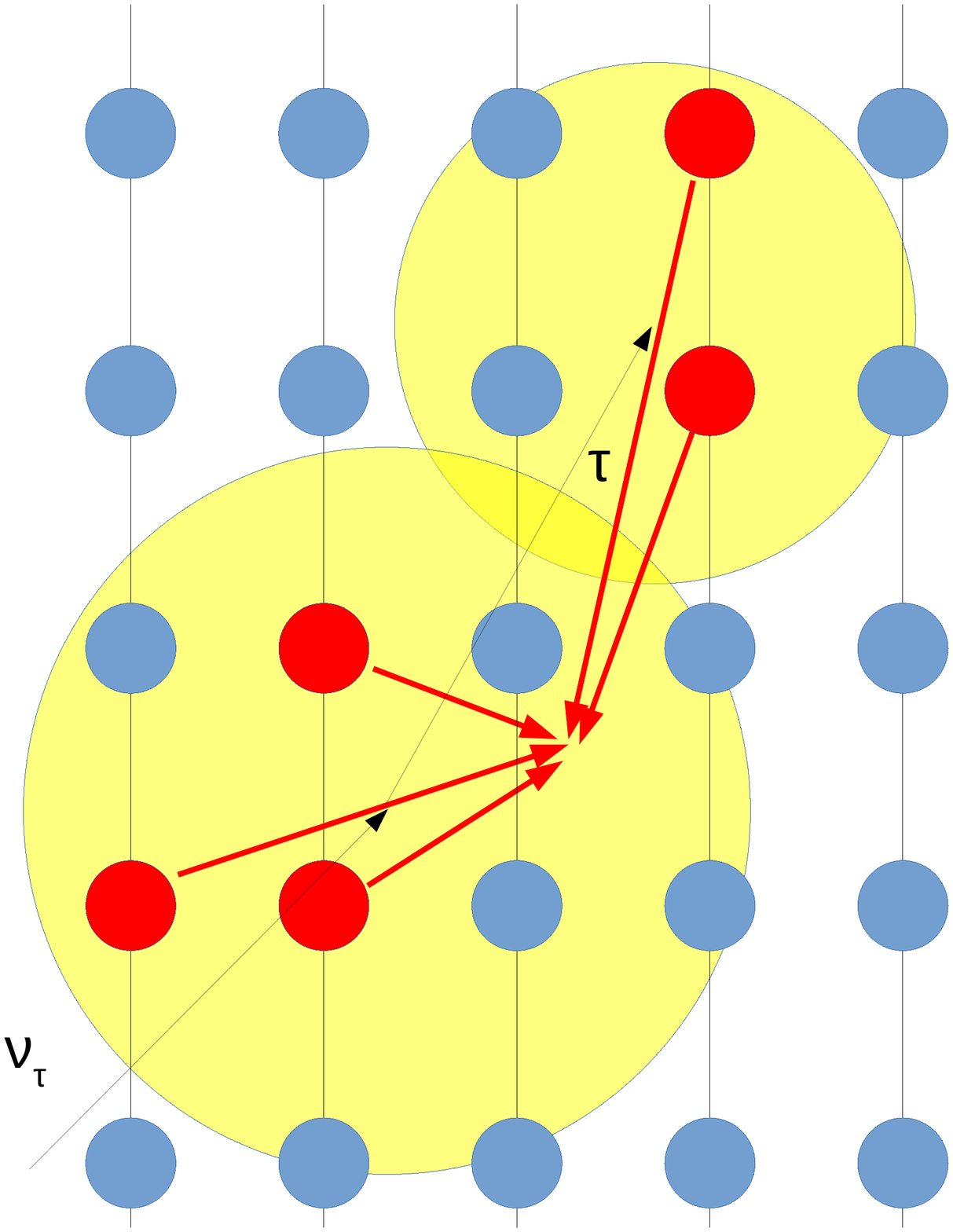}
			\label{obr:set_5_two_cascades}
		}        
	\end{multicols}
	\caption{Illustration of cascade position estimation. a) Pulses for position estimation correspond to one cascade only $\Rightarrow$ accurate position estimation. b) Pulses for position estimation correspond to two cascades $\Rightarrow$ inaccurate position estimation.} 
	\label{fig:set_5_illustration}
\end{figure}

Firstly, a set of five random pulses is chosen. These five pulses are used for cascade position and time estimation. The cascade position and time are estimated by solving a set of equations for distances between OMs that detected particular pulses and position of cascade vertex $d_{i}$:

\begin{equation}
d_{i} = \sqrt{(x-x_{i})^{2}+(y-y_{i})^{2}+(z-z_{i})^{2}} = c/n(t-t_{i}),
\label{eq:estimate_pos_time}
\end{equation}
where $x = (x,y,z,t)$ is space-time position of cascade vertex, and $x_{i} = (x_{i},y_{i},z_{i},t_{i})$ is space-time position of OM that recorded particular hit~\cite{Hartmann:Thesis:2006}. If these five pulses correspond to one cascade only, cascade position and time should be estimated accurately (see Fig.~\ref{obr:set_5_one_cascade}). If this set is composed of pulses from two cascades, then the cascade position and time are estimated inaccurately (see Fig.~\ref{obr:set_5_two_cascades}).

\begin{figure}[h!]
	\centering
	\begin{multicols}{2}
		\centering
		\subfloat[]{
			\includegraphics[width=\linewidth]{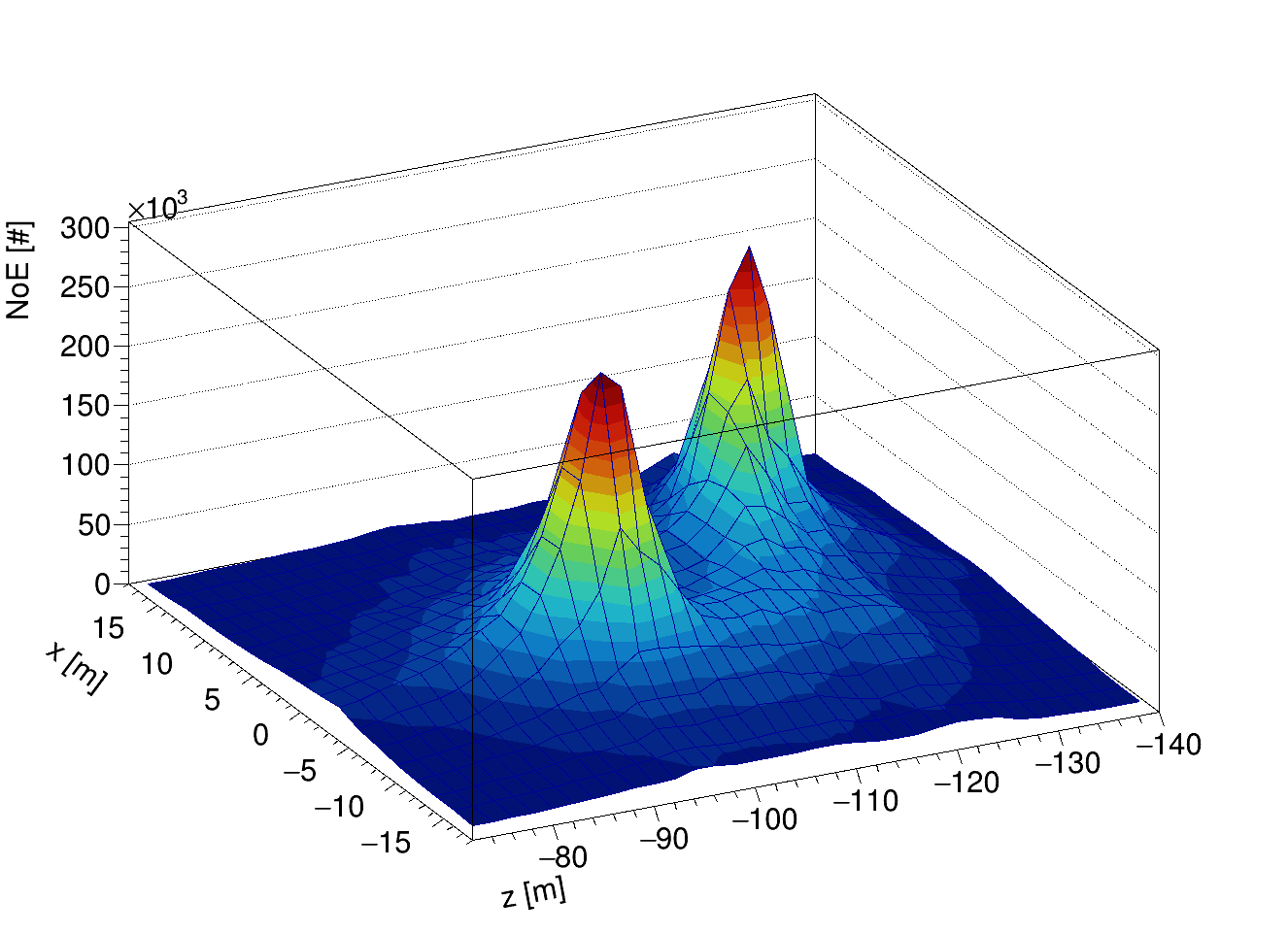} 
			\label{obr:XZ_plane}
		}
		\subfloat[]{
			\includegraphics[width=\linewidth]{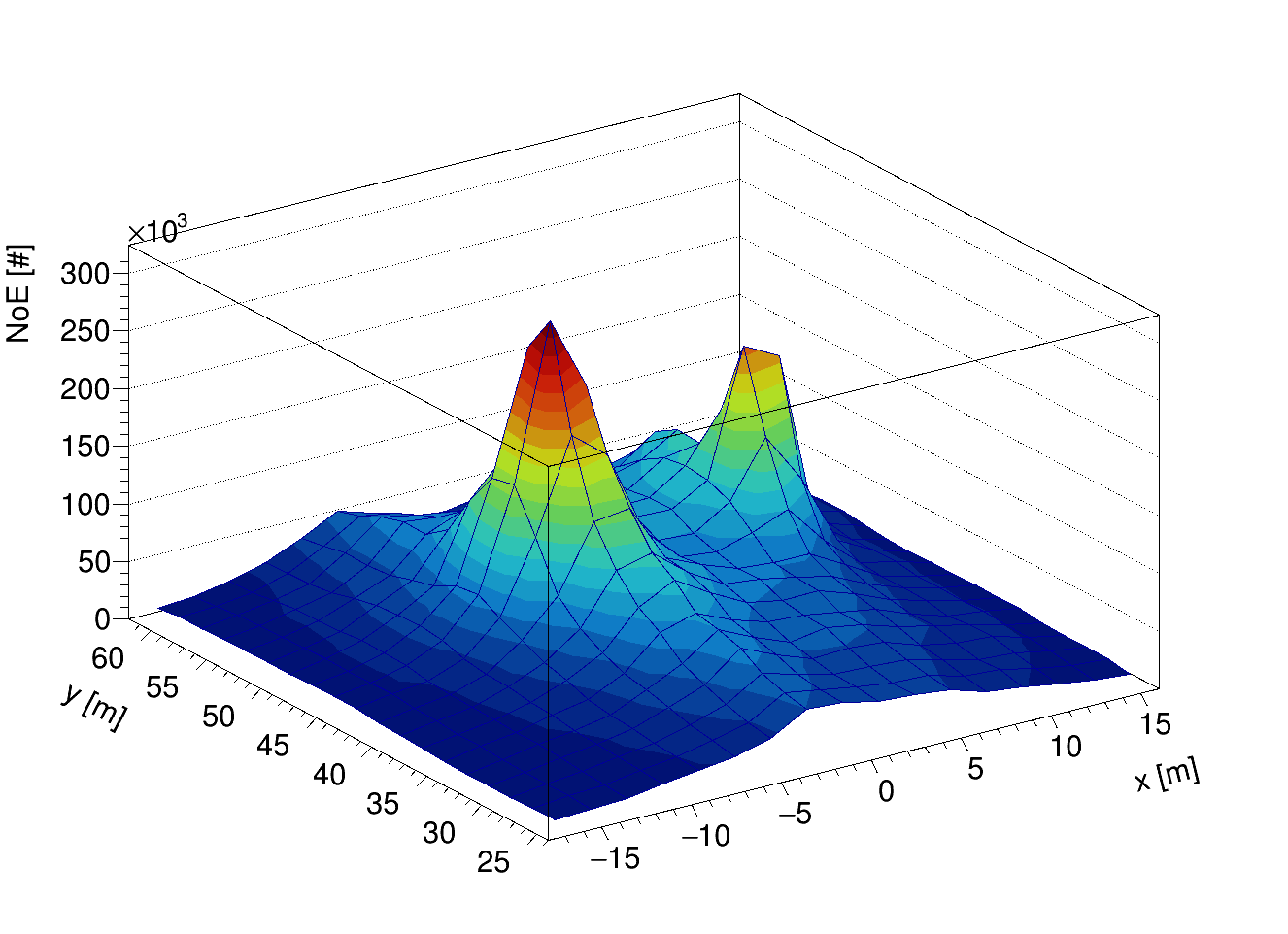}
			\label{obr:XY_plane}
		}        
	\end{multicols}
	\caption{The results of position estimation procedure, considering all combinations of five pulses, for MC $\nu_{\tau}$ event with $\nu_{\tau}$ cascade vertex position [9, 47, -126], $\tau$ cascade vertex position [-3, 45, -100]. The MC positions are expressed in meters. a) Display of XZ plane. a) Display of XY plane.}
	\label{fig:set_of_five_demonstration}
\end{figure}

 Demonstration of this method on one event, for all combinations of five pulses, is shown in Fig.~\ref{fig:set_of_five_demonstration}, two distinct peaks that correspond to vertices of two cascades are created. Accurately estimated position and time of cascade enable to select pulses that correspond to particular cascade. Pulses are selected with respect to the estimated position $\vec{R}$ and time $T$ of cascade according to the criterion: 

\begin{equation}
\mid T^{meas}_{i}-T^{exp}_{i}(\vec{R},T) \mid \lesssim \delta T,
\label{eq:tfilter}
\end{equation}
where $T^{meas}_{i}$ is detection time of the pulse, $T^{exp}_{i}$ is expected time of pulse detection, and $\delta T$ determines strictness of the criterion. There are also additional criteria on the charge of studied pulse, and the distance between cascade position and OM that detected studied pulse. The estimation of position and time of the cascade is computationally extensive if all possible combinations of five pulses are taken into account.  Therefore, if the chosen subset of pulses satisfies a set of certain conditions, this subset of pulses is chosen for single cascade reconstruction, and the selection of sets of five pulses is ceased.

The last step is to process two subsets of pulses with single cascade reconstruction~\cite{ICRC2019:cascades}. An event has to have at least 6 hits on at least 3 strings (6/3 filter) to be reconstructed with single cascade reconstruction. For that reason both subsets of pulses for double cascade reconstruction have to pass this criterion. Subsequently, selected sets of pulses undergo single cascade reconstruction procedure. In the single cascade reconstruction, time and position of the cascade are estimated by minimization of $\chi^{2}$:
\begin{equation}
\chi^2 = \frac{1}{N_{hit} - 4} \sum\limits_{i = 1}^{N_{hit}} \frac{ ( T_i^{meas} - T_i^{exp}(\vec{R},T))^2}{\sigma_t^2},
\label{eq:chi2}
\end{equation}
where $N_{hits}$ is number of hit OMs, $T_{i}^{meas}$ is experimentally measured time of pulse detected on $i^{th}$ OM, $T_{i}^{exp}$ is theoretically expected time of pulse detection on $i^{th}$ OM, given that cascade vertex time and position are $T$ and $\vec{R}$, and $\sigma_t$ is the uncertainty in time measurements. Energy and direction of a cascade are determined via minimization of likelihood function:
\begin{equation}
L =- \sum_{i=1}^{hit OM} log (P_{i}(q_{i} \mid Q_{i})) - \sum_{i = 1}^{unhitOM} log (P_{i}(q_{i} = 0 \mid Q_{i})),
\label{eq:likelihood}
\end{equation}
where $P_{i}$ is the Poisson probability of detecting charge $q_{i}$ on $i^{th}$ OM while charge $Q_{i}$ is expected. 

\begin{figure}[h!]
	\centering
	\begin{multicols}{2}
		\centering
		\subfloat[]{
			\includegraphics[width=\linewidth, height=4.5cm]{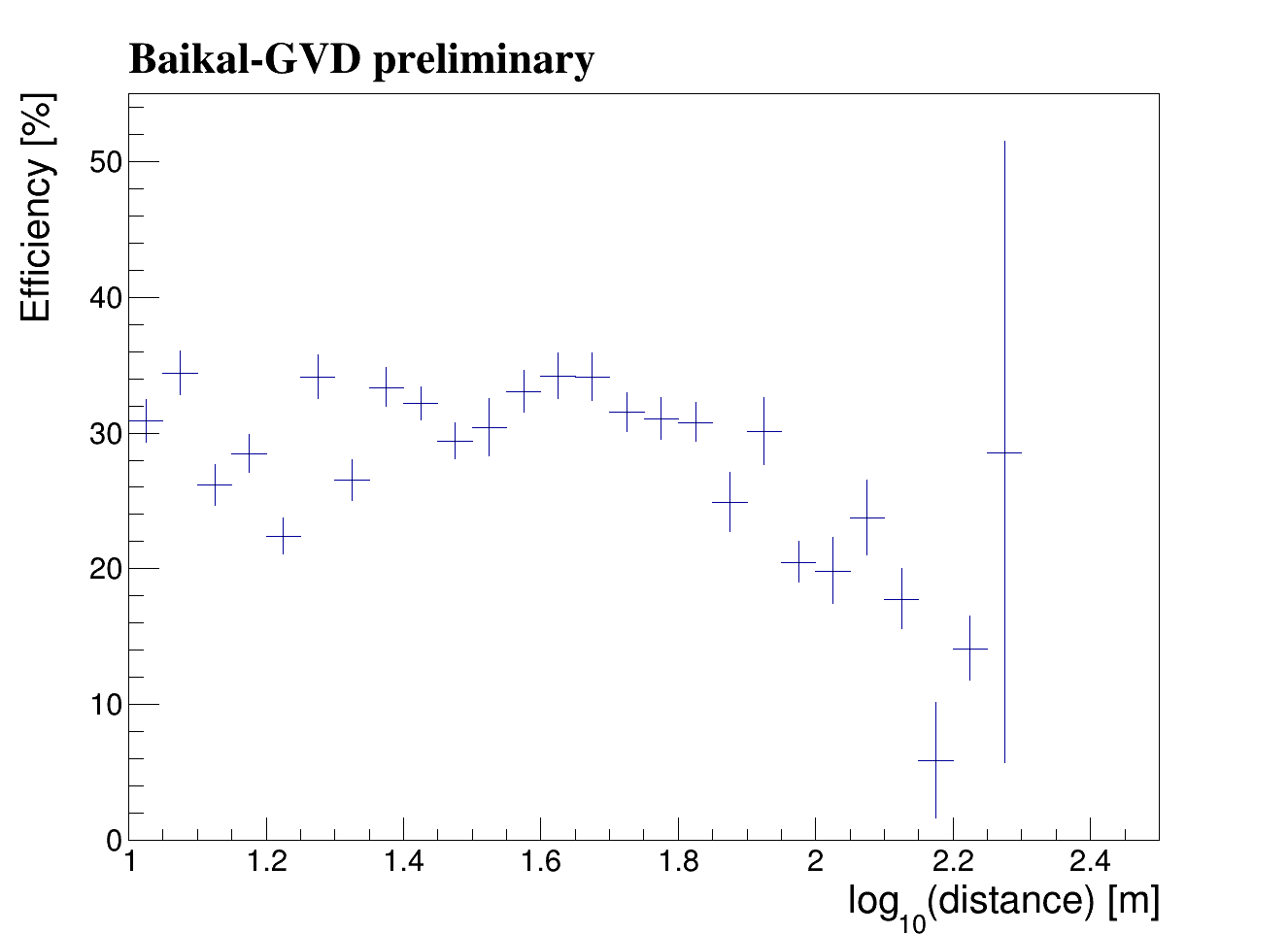}
			\label{obr:eff_63}
		}
		\subfloat[]{
			\includegraphics[width=\linewidth,height=4.5cm]{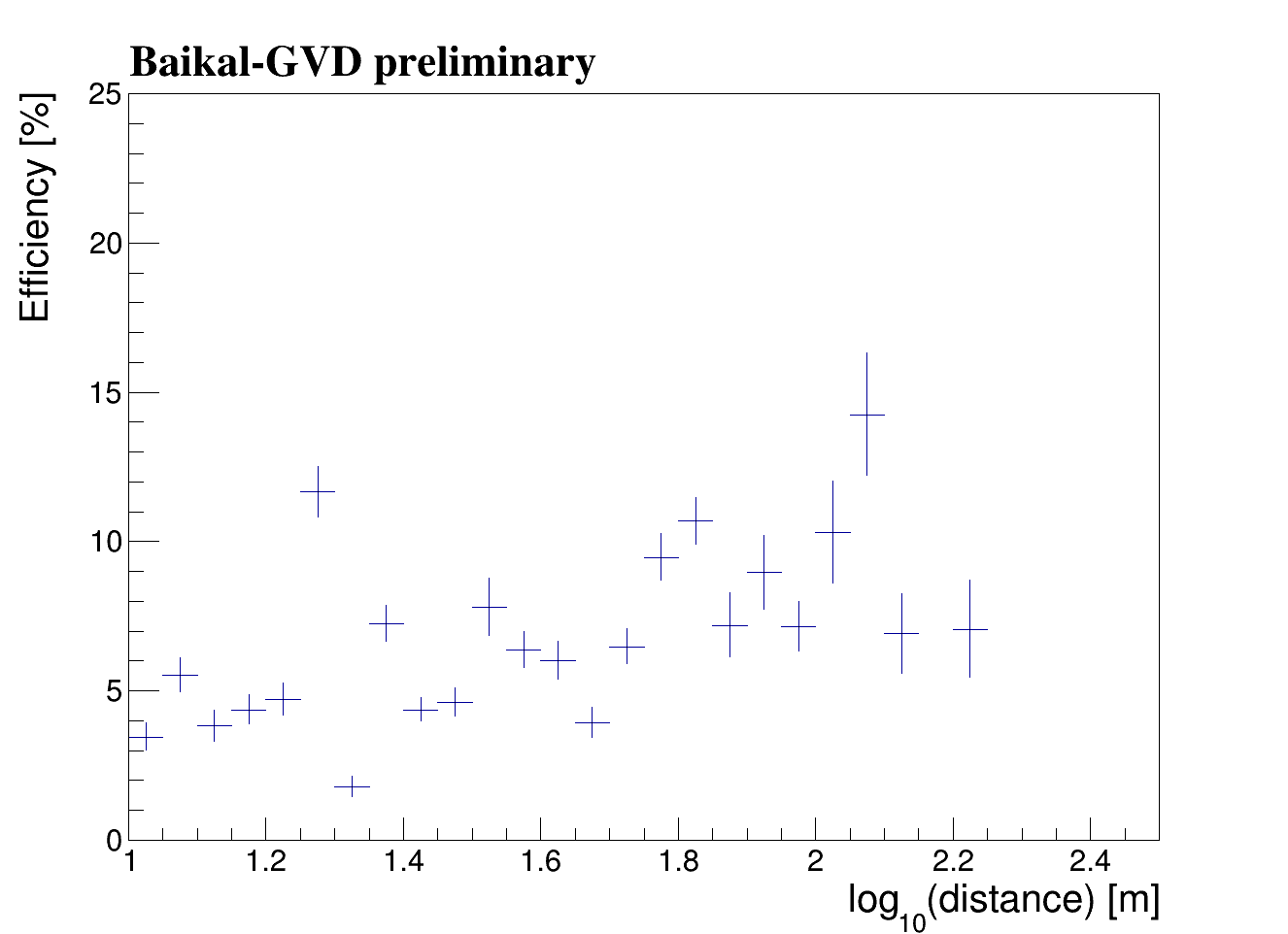}
			\label{obr:eff_reco}
		}        
	\end{multicols}
	\caption{a) The dependence of reconstruction efficiency on MC simulated distance between cascades after 6/3 filter. b) The distribution of reconstruction efficiency with respect to MC simulated distance between cascades after double cascade reconstruction (tolerance on mismatch between simulated and reconstructed position is 5~m).}
	\label{fig:eff_63_reco}
\end{figure}

\begin{figure}[h!]
	\begin{center}
		\includegraphics[width=0.925\textwidth]{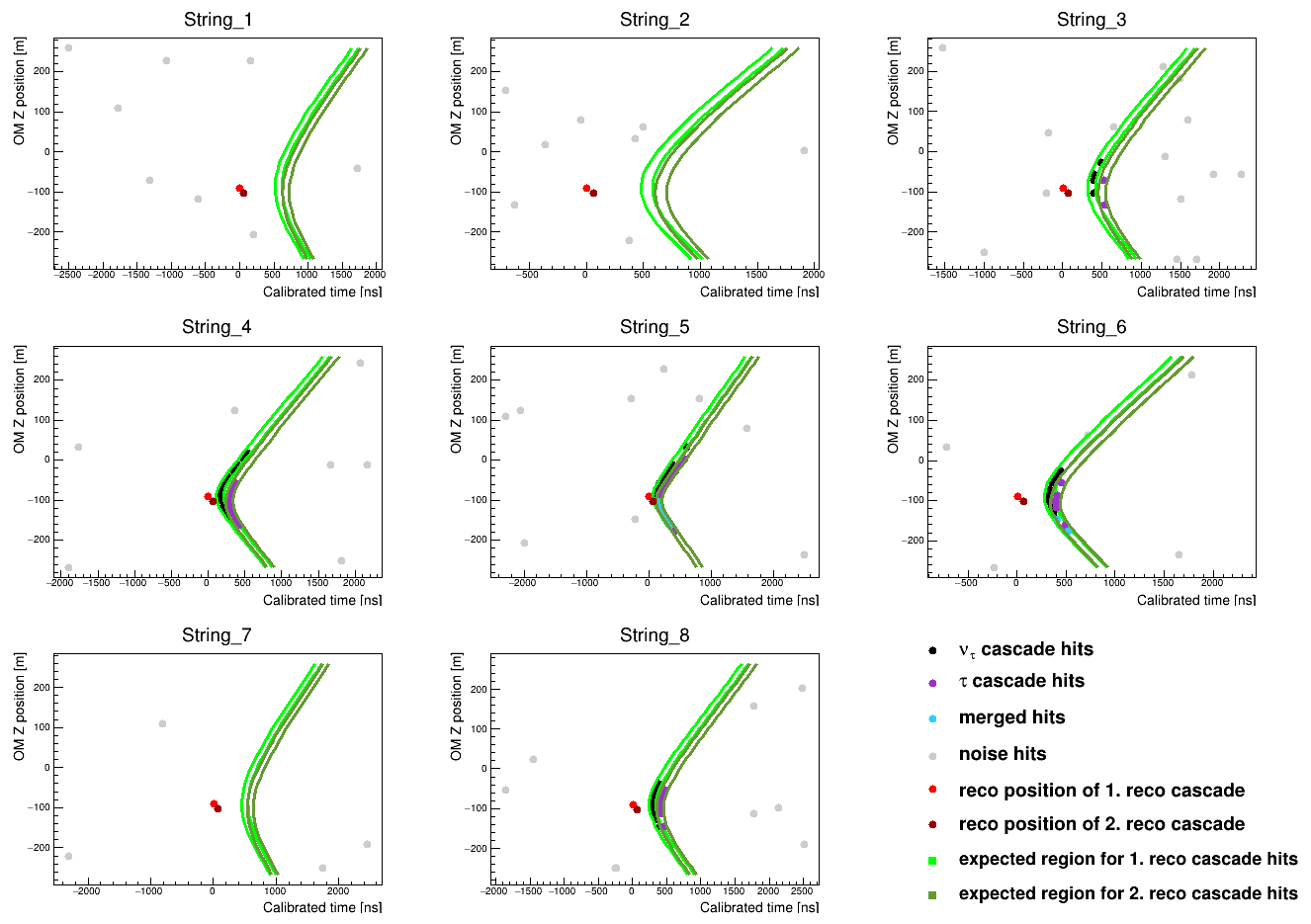}
		\caption{Reconstructed MC $\nu_{\tau}$ event visualization. The dependence of time on the Z coordinate for all eight strings is shown. The MC simulated distance between cascades is 18.17~m. The mismatch between simulated and reconstructed position of $\nu_{\tau}$ cascade is 1.84~m and 3.27~m for $\tau$ cascade.} 
		\label{fig:event_visualization}
	\end{center}%
\end{figure}

The Monte Carlo (MC) simulations of $\nu_{\tau}$ interaction were processed with aforementioned reconstruction algorithm~\cite{parallel}. Only double cascade events with distance between cascade vertices higher than 10~m were taken into account. 
The reconstruction efficiency of the double cascade events is defined as the ratio of events that passed certain criteria and total events simulated in MC. In Fig.~\ref{obr:eff_63}, the reconstruction efficiency after 6/3 filter is displayed. The reconstruction efficiency after whole double cascade reconstruction algorithm is shown in Fig.~\ref{obr:eff_reco} (tolerance on mismatch between simulated and reconstructed position is 5~m for both cascades). An example of a reconstructed double cascade event is displayed in Fig.~\ref{fig:event_visualization}.

\section{Multi-cluster Double Cascades Reconstruction}

The multi-cluster double cascade identification technique combines single cascade reconstruction algorithm with multi-cluster events. A double cascade event is tagged if two single cascades were reconstructed separately in two different clusters satisfying time coincidence condition.

Approximately 87\,000 cascades were reconstructed by means of single cascade reconstruction technique~\cite{parallel} in the experimental data collected in year 2019. The same dataset has been processed with multi-cluster double cascade identification algorithm. This method tagged one double cascade event which means that two single cascades were reconstructed within the coincidence time window. Reconstructed attributes of the two cascades are shown in Tab.~\ref{tab:Attributes}.

\begin{table}
	\caption{Reconstructed parameters of cascades from double cascade multi-cluster event: zenith angle, azimuth angle, reconstructed energy, number of hits used in reconstruction, and likelihood.}
	\label{tab:Attributes}
	\centering
	\scalebox{0.825}{%
		\begin{tabular}{c||c|c|c|c|c}
			& $\theta$ [rad] & $\phi$ [rad] & Energy [TeV] & nRecoHits & likelihood \\
			\hline \hline
			cascade 1 & 2.20 & 3.85 & 8.06 & 19 & 0.92
			\\ \hline
			cascade 2 & 2.32 & 3.66 & 4.72 & 17 & 1.08
			\\		
	\end{tabular}}
\end{table}
 
 According to the reconstructed positions of two cascades, the distance between cascades was determined as $\sim$~328.75~m. Taking into account reconstructed attributes of two cascades, the probability of $\nu_{\tau}$ origin of this event is close to zero.

\section{Conclusion}

The double pulse detection method based on machine learning technique BDT was developed. The signal efficiency gained is 99.6\%, and the background efficiency is 1.4\%. The usage of this method for multiple pulses identification and separation was studied.
The first single cluster double cascade reconstruction algorithm has been developed. The individual steps of this algorithm were described. The first preliminary results were studied.
The multi-cluster double cascade identification algorithm was applied on data collected in year 2019. One multi-cluster double cascade event was identified. According to the analysis of reconstructed parameters of two cascades, $\nu_{\tau}$ origin of this event can be excluded.

\section{Acknowledgments}
The work was partially supported by RFBR grant 20-02-00400. The CTU group acknowledges the support by European Regional Development Fund-Project No. CZ.02.1.01/0.0/0.0/16\_019/0000766. The CU group acknowledges the support by the VEGA Grant Agency of the Slovak Republic under Contract No. 1/0607/20 and support by the  Grants  of  Comenius  University  in  Bratislava, Slovakia UK/321/2021. We also acknowledge the technical support of JINR staff for the computing facilities (JINR cloud).

\end{document}